\documentclass{eptcs}
\usepackage{amsmath,amssymb}
\usepackage[all]{xy}

\def\shuf{\mathbin{\mathchoice
{\rule{.3pt}{1ex}\rule{.3em}{.3pt}\rule{.3pt}{1ex}
\rule{.3em}{.3pt}\rule{.3pt}{1ex}}%
{\rule{.3pt}{1ex}\rule{.3em}{.3pt}\rule{.3pt}{1ex}
\rule{.3em}{.3pt}\rule{.3pt}{1ex}}%
{\rule{.2pt}{.7ex}\rule{.2em}{.2pt}\rule{.2pt}{.7ex}
\rule{.2em}{.2pt}\rule{.2pt}{.7ex}}%
{\rule{.3pt}{1ex}\rule{.3em}{.3pt}\rule{.3pt}{1ex}
\rule{.3em}{.3pt}\rule{.3pt}{1ex}}%
}}

\def\raa{\longrightarrow}
\newcommand{\Nat}{\mathbb{N}}
\def\ra{\rightarrow}

\def\ss{\subseteq}

\def\raa{\longrightarrow}
\newcommand{\ew}{\lambda}

\begin{document}

\title{On the Existence of Universal Finite or Pushdown Automata
}
\def\titlerunning{On the Existence of Universal Finite and Pushdown Automata}
\author{Manfred Kudlek\footnote{
Sadly, Manfred Kudlek passed away June 18, 2012, before the publication of these proceedings}
\institute{Fachbereich Informatik, MIN-Fakult\"at,
 Universit\"at Hamburg, DE}
}
\def\authorrunning{Kudlek}
\date{}
\maketitle


\begin{abstract}
We investigate the (non)-existence of universal automata for some classes
of automata, such as finite automata and pushdown automata, and
 in particular the influence of the re�presentation and encoding function.
An alternative approach, using transition systems, is presented too.
\end{abstract}

\section{Introduction}

It is well known that there exist universal Turing machines (UTM).
Such a UTM simulates any special Turing machine (TM) $M$ in a certain
way. There are several ways of simulation. One is that a UTM $U$ 
simulating a TM $M$ with input $w$ halts if and only if $M$ halts
on input $w$. Another possibility is that any computation step of $M$
is simulated by $U$ using some number of steps which are be restricted
by some complexity function. In very small UTM's this can be
exponential.

Almost all UTM's constructed so far are deterministic, simulating
deterministic TM's. In \cite{KM} it has been shown that there exist UTM's simulating
all special TM's with complexity constraints. These complexity constraints,
for space or time, are from a subclass of all primitive recursive functions
over one variable. The UTM's have the same complexity constraints.

In both cases, general TM's and those with complexity constraint,
the specific TM $M$ and its input $w\in\Sigma(M)^*$, where $\Sigma(M)$ is
the alphabet of $M$, have to be encoded. Such an encoding, and also the
decoding, can be achieved by deterministic finite state transducers (DFST),
which means that encoding and decoding is bijective. The input for a
UTM $U$, to simulate $M$ with input $w$, can then have the form
$c_m(M){\#}c_i(w)$ where $c_m$, $c_i$ are the encoding functions for $M$, $w$,
respectively.

If one intends to construct universal machines for weaker automata classes it
should be kept in mind that encoding and decoding for such automata should not
exceed the power of deterministic versions of those machines. Otherwise too
much power and information could be hidden in the encoding.

In \cite{Kud} it has been shown, under this condition, that there don't exist universal
1-way finite automata (FA),  neither deterministic (DFA) nor nondeterministic (NFA) ones. The proof uses arguments on the
number of states of such automata.

So the question arises whether there exist universal universal pushdown automata (UPDA), and if so if
encoding and decoding can be achieved by DFST's, or if deterministic pushdown
transducers (DPDT) are necessary. For general TM's DFST's suffice for encoding
and decoding (e.g. \cite{Min}).

\section{Transducers}

 Pushdown transducers (see e.g. \cite{Gin,Gur}) are just the analogon
to finite state transducers, i.e. pushdown automata with output.

\bigskip\noindent
Formally:

A {\it (non-deterministic) pushdown transducer (PDT)} is a construct
$(Q,\Sigma_i,\Sigma_o,\Delta,{\$},q_0,Q_f,\rho)$ where
%
\begin{center}
\begin{tabular}{ll}
$Q$ & set of states \\
$\Sigma_i$ & input alphabet \\
$\Sigma_o$ & output alphabet \\
$\Delta$ & pushdown alphabet \\
${\$}\in\Delta$ & stack bottom symbol \\
$q_o\in Q$ & initial state \\
$Q_f\subseteq Q$ & set of final states \\
$\rho\subseteq Q\times\Sigma_i^*\times\Sigma_o^*\times\Delta^*
\times\Delta^*\times Q$ &
the transition relation\ . \\
\end{tabular}
\end{center}

\bigskip

A {\it deterministic pushdown transducer (DPDT)} is the deterministic version
of PDT. That is, $\rho$ is a function
$\rho:Q\times\Sigma_i^*\times\Delta^*\ra\Sigma_o^*\times Q$.

In the sequel a normal form of PDT's will be considered, being quasi lettering
in input and pushdown, i.e.
$\rho\subseteq Q\times(\Sigma_i\cup\{\lambda\})\times(\Sigma_o\cup\{\lambda\})
\times(\Delta\cup\{\lambda\})\times(\Delta\cup\{\lambda\})\times Q$.

\bigskip

Contrary to regular languages ({\bf REG}) which are closed under finite state
transductions the context-free languages ({\bf CF})
are not closed under pushdown transductions, neither non-deterministic nor
deterministic. Even linear context-free languages give sets outside {\bf CF}
if a deterministic pushdown transduction is applied. This is shown by the
following examples of DPDT's where $\tau$ denotes the function
defined by a deterministic transducer.

\medskip

\noindent
Let $L_1 = \{0^n1\ |\ n\geq 0\}\in {\bf REG}$ and

$T_1=(\{q_0,q_1\},\{0,1\},\{0,1\},\{q_0\},\{q_1\},\rho_1)$
with

$\rho_1 = \{(q_0,0,0,{\$},0{\$},q_0),(q_0,0,0,0,00,q_0),
(q_0,1,\lambda,{\$},{\$},q_2),$

\hspace{1cm}$(q_0,1,\lambda,0,0,q_1),
(q_1,\lambda,1,0,\lambda,q_1),(q_1,\lambda,\lambda,{\$},{\$},q_2)\}$.

Then $\tau(L_1) = \{0^n1^n\ |\ n\geq 0\} \not\in {\bf REG}$.

\medskip

\noindent
Let $L_2 = \{0^n1^n0\ |\ n>0\}\in {\bf LIN}$ and

$T_2=(\{q_0,q_1,q_2,q_3\},\{0,1\},\{0,1\},\{q_0\},\{q_3\},\rho_2)$
with

$\rho_2=\{(q_0,0,0,{\$},0{\$},q_0),(q_0,0,0,0,00,q_0),
(q_0,1,1,0,10,q_1),$

\hspace{1cm}$(q_1,1,1,1,11,q_1),
(q_1,0,\lambda,1,1,q_2),(q_2,\lambda,0,1,\lambda,q_2),$

\hspace{1cm}$(q_2,\lambda,1,0,\lambda,q_2),
(q_2,\lambda,\lambda,{\$},{\$},q_3)\}$.

Then $\tau(L_2) = \{0^n1^n0^n1^n\ |\ n\geq 0\} \not\in {\bf CF}$.

\medskip

\noindent
Let $L_3 = \{wcw^Rc\ |\ w\in\{0,1\}^+\}\in {\bf LIN}$ and

$T_3=(\{q_0,q_1,q_2,q_3\},\{0,1\},\{0,1\},\{q_0\},\{q_3\},\rho_3)$
with

$\rho_3=\{(q_0,0,0,{\$},0{\$},q_0),(q_0,1,1,{\$},1{\$},q_0),(q_0,0,0,0,00,q_0),$

\hspace{1cm}$(q_0,0,0,1,01,q_0),(q_0,1,1,0,10,q_0),(q_0,1,1,1,11,q_0),$

\hspace{1cm}$(q_0,c,c,0,0,q_1),(q_0,c,c,1,1,q_1),(q_1,0,0,{\$},0{\$},q_1),$

\hspace{1cm}$(q_1,1,1,{\$},1{\$},q_1),(q_1,0,0,0,00,q_1),
(q_1,0,0,1,01,q_1),$

\hspace{1cm}$(q_1,1,1,0,10,q_1),(q_1,1,1,1,11,q_1),(q_1,c,c,0,0,q_2),$

\hspace{1cm}$(q_1,c,c,1,1,q_2),
(q_2,\lambda,\lambda,0,0,q_2),(q_2,\lambda,\lambda,1,1,q_2),$

\hspace{1cm}$(q_2,\lambda,\lambda,c,c,q_2),(q_2,\lambda,\lambda,{\$},{\$},q_3)\}$.

Then $\tau(L_3) = \{wcw^Rcwcw^Rc\ |\ w\in\{0,1\}^+\} \not\in {\bf CF}$.

\bigskip

One also might consider 2-way FST's (2FST). However, {\bf REG} is not closed
under 2-way finite state transductions, as can be seen from the following
example of a 2DFST ($L,R,M$ denote move to left, right, or not,
respectively).

\noindent
Let $L_4=\{awb\ |\ w\in\{0,1\}^*\}\in {\bf REG}$ and

$T_4=(\{q_0,q_1,q_2,q_3\},\{0,1,a,b\},\{0,1,a\},\{q_0\},\{q_4\},\rho_2)$
with

$\rho_4=\{(q_0,a,a,R,q_0),(q_0,0,0,R,q_0),(q_0,1,1,R,q_0),(q_0,b,\lambda,L,q_1),$

\hspace{1cm}$(q_1,0,\lambda,L,q_1),(q_1,1,\lambda,L,q_1),(q_1,a,a,R,q_2),$

\hspace{1cm}$(q_2,0,0,R,q_2),(q_2,1,1,R,q_2),(q_2,b,\lambda,M,q_3)$.

Then $\tau(L_4)=\{awaw\ |\ w\in\{0,1\}^*\}\not\in {\bf CF}$.

\bigskip

\section{Representations}

A representation of FA, PDA, TM's etc. has to contain information on the
set of states, initial and final states, alphabets, and the set of transitions:

$(Q,\Sigma,\Delta,Q_0,Q_f,R)$ with e.g. 
$R\subseteq Q\times\Sigma^*\times\Delta^*\times\Delta^*\times Q$ for a PDA.

\noindent
Usually $R$ is represented by an ordered list of elements from $R$, together with
lists for $Q_0$ and $Q_f$. For non-deterministic machines one might also allow
repetitions of list elements. This can give a regular set of representations if
$Q$, $\Sigma$, and $\Delta$ are fixed.

FA and PDA usually are represented by the list $R$ of their transitions,
putting together the tuples $(q,x,q')$ or $(q,x,y,y',q')$ for FA or PDA
respectively, where $x\in(\Sigma\cup\{\lambda\})$, and
$y,y'\in(\Delta\cup\{\lambda\})$.

For FA one has a representation $R\in(Q\cdot (\Sigma\cup\{*\})\cdot Q)^+$, and for PDA
$R\in(Q\cdot (\Sigma\cup\{*\})\cdot (\Delta\cup\{*\})\cdot
 (\Delta\cup\{*\})\cdot Q)^+$
where $*$ stands for $\lambda$.
Together with the input $w$ this gives a representation $R(M){\#}w$.
But one might think also of a representation $w{\#}R(M)$ or even
$R(M)\shuf {\#}\shuf w$ where $\shuf$ is the shuffle operation,
and ${\#}$ a separation symbol.

If the first version of representation is encoded by a DFST $T$ the result is a word
$\tau(R(M))\tau({\#})\tau(w)$ where $\tau$ is the function associated to $T$. 
This follows from the fact that $T$ is working 1-way.
The same conditions hold for the `inverse', namely for a DPDT $T'$ with
associated function $\tau'$. Furthermore,
$\tau'(\tau(R(M))\tau({\#})\tau(w)) = R(M){\#}w$ has to hold.

\medskip

Considering representations of FA with arbitrary sets of states $Q$ and arbitrary
alphabets $\Sigma$, states $q_k$ can be represented by $qa^k$ ($0<k$) and symbols $x_m$ by
$xa^m$ ($0<m$) over the finite alphabet $\{q,x,a\}$. Then the set of all
representations of finite automata is given by

$F=(\{q\}\cdot \{a\}^+\cdot (\{x\}\cdot\{a\}^+\cup\{*\})\cdot \{q\}\cdot\{a\}^+)^+$.

Clearly, $F\in {\bf REG}$. But note that this holds only for non-deterministic FA.
In the case of DFA there is the condition that a pair $(q_k,x_m)$
of state and symbol, represented by $qa^k$ and $xa^m$,
can appear only once as first components in the list of transitions. 

An analogous property also holds for PDA.

\medskip

Since {\bf REG} and {\bf CF} are closed under FST mappings, encoding (and
decoding) will not lead out from these classes. However, the examples
above show that this does not hold for PDT mappings. Therefore, not to gain too
much power it might be reasonable to have a condition that those PDT mappings
used for encoding and decoding are not leading out from the class {\bf CF}.

\section{Universality?}

In the sequel we shall consider PDA $M$ accepting languages $L(M)\subseteq\{0,1\}^*$.
Assume that there exists a universal PDA $U$ simulating all specific PDA $M$ over
$\{0,1\}$. Denote this class by ${\bf CF}_2$.
Let $\Sigma_U$ be the alphabet of $U$. Then, with the first version of
representation,

$L(U) = \{\tau(R(M))\tau({\#})\tau(w)\ |\ L(M)\in {\mathbf C}{\mathbf F}_2,
 w\in L(M)\} \in {\bf CF}$.

Consider now the special regular (context-free) languages $\{w\}\subset\{0,1\}^*$.
A representation of a DFA $M$, being also a PDA or 2DFA, accepting exactly
the language $\{w\}$,
e.g. looks like
\mbox{$R(M)=q_0wq_1=\varphi(w)$}.
Together with an input $v\in\{0,1\}^*$ the representation has the form
$R(M){\#}v=q_0wq_1{\#}v$.
A DFST maps this into $\tau(\varphi(w))\tau({\#})\tau(v)$.
Clearly,

$S=\{\tau(\varphi(w))\tau({\#})\tau(v)\ |\ v,w\in\{0,1\}^*\}$

\hspace{.5cm}$=\{\tau(\varphi(w))\ |\ w\in\{0,1\}^*\}\cdot\{\tau({\#})\}\cdot
\tau(\{0,1\}^*)
\in {\bf REG}$.

Now 

$L(U)\cap S = \{\tau(\varphi(w))\tau{\#}\tau(v)\ |\
w,v \in\{0,1\}^*, v=w\}$

\hspace{1cm}$=\{\tau(\varphi(w))\tau({\#})\tau(w)\ |\ w\in\{0,1\}^*\} \in {\bf CF}$

since {\bf CF} is closed under intersection with regular sets.

Applying the `inverse' DFST mapping $\tau'$ yields

$\tau'(L(U)\cap S) = \{\varphi(w){\#}w\ |\ w\in\{0,1\}^*\}\in {\bf CF}$.

Another DFST mapping $\psi$ with $\psi(q_0)=\psi(q_1)=\lambda$, $\psi(0)=0$,
$\psi(1)=1$ gives

$\psi(\tau'(L(U)\cap S)) = \{w{\#}w\ |\ w\in\{0,1\}^*\}\in {\bf CF}$,

a contradiction.

\bigskip

These considerations can be summarized as

\medskip

\noindent
{\bf Theorem 1:} If encoding and decoding of specific finite or pushdown automata
have to be
achieved by DFST then there doesn't exist a universal finite automaton,
or
2-way finite automaton or pushdown automaton, simulating all specific
finite automata.
\hfill$\Box$

\bigskip

It should be remarked, however, that the proof of this theorem cannot be used to
show that the statement also holds for all quasi lettering finite automata.
The reason is that the DFST for encoding has to know the length of $w$ for the
representation $q_0x_1q_1\cdots x_kq_k$ where $x_i\in\{0,1\}$ and all $q_i$
are different.

Therefore we give another proof that this theorem also holds for quasi-lettering
automata. For non-deterministic (quasi-lettering) FA it can be assumed that there
is exactly one initial and exactly one final state, and that the first element
in the list has the form $(q_1,x,q)$ where $q_1$ is the initial state and
$x\in\Sigma$, and that the last element has the form $(q_2,\lambda,q_2)$ where
$q_2$ is the final state.

\medskip

Let
$M_{\mu}=(Q,\{0,1\},q_1,\{q_2\},R_{\mu})$
 a FA, illustrated in Figure 1, where
 
\noindent 
  $\mu=(i,i',i'',j,j',j'',k,k',k'',\ell,\ell',\ell'')$ and

\noindent
  $2<i,i',i'',j,j',j'',k,k',k'',\ell,\ell',\ell''\leq |Q|$
are fixed and







$R(M_{\mu})=\{(q_1,0,q_i),(q_1,1,q_k),(q_{i'},0,q_{j'}),(q_{i''},1,q_{\ell''}),$

\hspace{1.2cm}$(q_{k''},0,q_{j''}),(q_{k'},1,q_{\ell'}),(q_j,0,q_2),(q_{\ell},1,q_2),(q_2,\lambda,q_2)\}$.

	\begin{figure}[htbp]
		\centerline{
		$\xymatrix@R=5pt@C=20pt{
			&& *+[o][F]{i} 
			& *+[o][F]{i'} 
				\ar[rrr]^{0}
			&&& *+[o][F]{j'} 
			& *+[o][F]{j}
				\ar[ddrr]^{0}\\
			&&& *+[o][F]{i''} 
				\ar[ddrrr]^<<<<{1}|{\hole}
			&&& *+[o][F]{j''} \\
			*++[o][F]{1} 
				\ar[uurr]^{0}
				\ar[ddrr]^{1}
			&&&&&&&&&*++[o][F]{2} \\
			&&& *+[o][F]{k''} 
				\ar[uurrr]^<<<<{0}
			&&& *+[o][F]{l''} \\
			&& *+[o][F]{k} 
			& *+[o][F]{k'} 
				\ar[rrr]^{1}
			&&& *+[o][F]{l'} 
			& *+[o][F]{l}
				\ar[uurr]^{1}\\
		}$
		}
		\caption{FA}
		\label{fig:upda2}
	\end{figure}

$L(M_{\mu})=\{000,011,100,111\}$ implies $i=i'=i''$, $j=j'=j''$, $k=k'=k''$,
$\ell=\ell'=\ell''$.
Note that $M_{\mu}$ is also a PDA.

An encoding of $R(M_{\mu})$ is given e.g. by

$\tau(R(M_{\mu}))=qaxbqa^i\cdot qaxcqa^k\cdot qa^{i'}xbqa^{j'}\cdot qa^{i''}xcqa^{\ell''}\cdot
$

\hspace{1.5cm}$qa^{k''}xbqa^{j''}\cdot qa^{k'}xcqa^{\ell'}\cdot qa^jxbqa^2\cdot qa^{\ell}xcqa^2\cdot
 qa^2*qa^2$


$\tau$ is a DFST mapping.
Note that $\tau$ actually depends on $Q$.

\medskip

Now assume that

$U=\{\tau(R(M))\tau({\#})\ |\ M\in {\rm QLFA}, \{000,011,100,111\}\subseteq L(M)\}$

\hspace{.8cm}$\cdot\tau(\{000,011,100,111\})$

is context-free, where QLFA denotes the class of quasi-lettering FA with
set of states $Q$, initial state $q_1$, final state $q_2$, and alphabet $\{0,1\}$.

Define
$S=\{\tau(R(M_{\mu}))\tau({\#})\ |\ 2<i,i',i'',j,j',j'',k,k',k'',\ell,\ell',\ell''\leq |Q|\}$

\hspace{2cm}$\cdot\tau(\{000,011,100,111\})$.

Obviously, $S$ is regular, implying that $U\cap S$ is context-free. But

$U\cap S = \{qaxbqa^i\cdot qaxcq^k\cdot qa^ixba^{\ell}\cdot q^ixcqa^{\ell}\cdot
qa^kxbqa^j\cdot qa^kxcqa^{\ell}\cdot qa^jxbqa^2$

\hspace{1.2cm}$\cdot qa^{\ell}xcqa^2\cdot qa^2*qa^2\ |\ 2<i,j,k,\ell\leq |Q|\}$

is not context-free, a contradiction.

\bigskip

If 2-way pushdown transducers are allowed however, and the encoding is not required to
have the form $\tau(R(M)\tau(\#)\tau(w)$,
e.g. in a more general form $\tau(R(M){\#}w)$,
 it is possible to construct a universal PDA,
following an idea by Gh. P\u{a}un et. al.. To be more precise, there exists a
quasi-lettering universal PDA simulating all lettering PDA
in this way.

\medskip
Let $M=(Q,\Sigma,\Delta,\delta,q_0,\{q_f\})$ with $\Sigma=\{0,1\}$
be a lettering PDA, and 
$R(M)\subseteq \{Q\times \Sigma\times
\Delta\times \Delta\times Q\}^* $
be its representation.
Suppose we allow $\tau(R(M)\# w)$ to be transduced by a 2-way PDT
 (2PDT).
Then one can choose the coding
$\tau(R(M)\# x_1x_2\dots x_n)=x_1 \sigma(R(M)) x_2 \sigma(R(M))
 \dots x_n \% \sigma(R(M))$
where $x_i\in\Sigma$, and $\sigma(R(M))$ is a coding of $R(M)$ in a fixed alphabet.
Note that $\sigma$ depends on $Q$. 
For simplicity let $\sigma=\iota$ at first where $\iota$ is the
identity function. Then $\tau$ can be 
calculated by a 2PDT that works as follows:
\begin{enumerate}

	\item First go right until the place after \# and print the first symbol of $w$.

	\item Go left and push $a\in\Delta$ to the stack until reading \#.

	\item Go left to the beginning of the input. 

	\item Print $R(M)$ until reading \#.
	
	\item Go further right and simultaniously pop $a\in\Delta$ from the stack.
On empty stack check if the end of the input has been reached.
	
	\begin{enumerate}
		\item If so, print `\%', go to the beginning of the input and copy $R(M)$ one last time and halt.
		\item Otherwise print the current input symbol and repeat from step 2.
	\end{enumerate}
\end{enumerate}

 Consider the following PDA $U$ (Figure 2) with 
		  $\Sigma_U=\{0,1\}\cup (Q\times\Sigma\times\Delta\times\Delta\times\Sigma)$ and
		 \mbox{ $\Delta_U=\{a\}\cup Q$} for all possible $Q,\Sigma,\Delta$.
		  The $\Sigma_U,\Delta_U$ are not finite, but when chosen an appropiate
		  encoding $\sigma$ for the representation of simulated automata,
		  the following idea for an UPDA works with finite alphabets 
		  $\Sigma_U,\Delta_{U}$. For now, assume any transition $(q,x,a,b,q')$
		  in the simulated automaton to be an atomic symbol of our UPDA.
		  
		  At any time $U$'s stack consists of
		  a word $q\delta_m\delta_{m-1}\dots\delta_0$ where
                  $\delta=\delta_m\dots\delta_0$ is the stack content
                  and $q$ the current state
		  of $M$ during a simulation. In state $R$, $U$ first reads a symbol
		  $x_i$ from the input (word $w$) and afterwards checks whether the 
		  simulated PDA could have read $x_i$ by travelling through the input
		  $R(M)$ and looking for a transition $(q,x_i,\delta_r,\delta_w,q')$ with
		  $q$ being the state $M$ is currently in, and storing the new state
		  and changed stack content of $M$ in its own stack.

	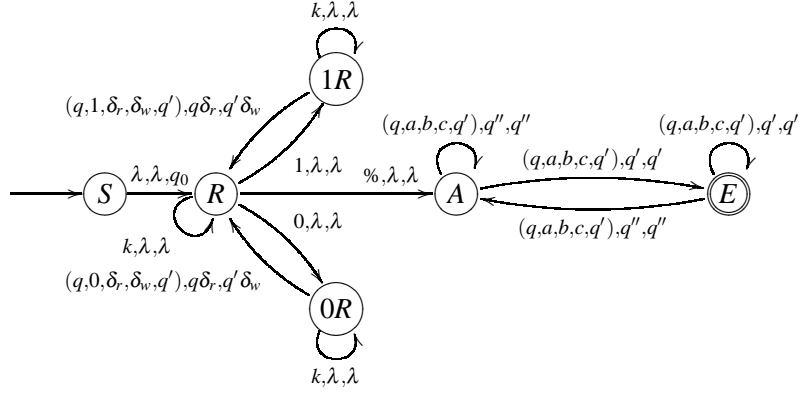
\begin{figure}[htbp]
		\centerline{
		$\xymatrix{
			&&& *++[o][F]{1R} 
				\ar@(ul,ur)[]^{k,\lambda,\lambda}
				\ar@/_/[dl]_{(q,1,\delta_r,\delta_w,q'),q\delta_r,q'\delta_w} \\
				*{} \ar[r]
			& *++[o][F]{S} \ar[r]^{\lambda,\lambda,q_0} 
				& *++[o][F]{R} 
					\ar@(l,d)[]_{k,\lambda,\lambda}
					\ar@/_/[ur]_{1,\lambda,\lambda} 
					\ar@/^/[dr]^{0,\lambda,\lambda}
					\ar[rr]^>>>>>>{\%,\lambda,\lambda}
				&& *++[o][F]{A} 
					\ar@(ul,ur)[]^{(q,a,b,c,q'),q'',q''}
					\ar@/^/[rrr]^{(q,a,b,c,q'),q',q'}
				&&& *++[o][F=]{E} 
					\ar@(ul,ur)[]^{(q,a,b,c,q'),q',q'}
					\ar@/^/[lll]^{(q,a,b,c,q'),q'',q''}\\
			&&& *++[o][F]{0R}
				\ar@(dl,dr)[]_{k,\lambda,\lambda}
				\ar@/^/[ul]^{(q,0,\delta_r,\delta_w,q'),q\delta_r,q'\delta_w} \\
		}$
		}
		\caption{Universal PDA}
		\label{fig:upda}
	\end{figure}

In the construction we have used $(q,x,a,b,q')$ etc. as one symbol,
but it works also if an encoding $\sigma$ over a finite alphabet $\Sigma_U$ is used.
E.g., $\sigma((q_i,x,\delta_j,\delta_k,q_{\ell}))=DSA^ixPB^jPB^kSA^{\ell}$
with $x\in\Sigma=\{0,1\}$ gives such an encoding.
 Then, point 4\ in the transduction $\tau$ would contain 
additional steps to encode a transition. E.g. to read symbol $q_i\in Q$
and print $Sa^i$ and so on.
Also, the UPDA $U$ must have additional components for decoding
in all states that have outgoing transitions reading a tuple $(q,x,a,b,q')$
from the input.

This can be achieved as follows. Let $\Sigma_U=\{A,B,D,P,S,T,0,1\}$ where
$D,P,S,T$ are markers. $D$ is the deliminator of
$DSA^iyPB^jPB^kSA^{\ell}$ encoding $(q_i,y,\delta_j,\delta_k,q_{\ell})$,
and $T$ the deliminator of a block $\sigma(R(M))$.

In a step $U$, after reading $x\in\{0,1\}$, goes into different states according to $x$.
The stack of $U$ contains $SA^mPB^nPB^r\cdots \$$ where $\$$ is the bottom symbol.
$U$ non-determistically goes to some $DSA^iyPB^jPB^kSA^{\ell}$ within $\sigma(R(M))$.
If $i=m$, $x=y$, and $j=n$ then $SA^mPB^n\cdots \$$ in the stack is
replaced by $SA^{\ell}PB^k\cdots \$$. Otherwise, $U$ goes into a sink.
Note that $U$ is a quasi-lettering PDA.

With a slight modification in the encoding it can be shown that the result also holds
for PDA $M$ not being lettering for the pushdown alphabet
$\Delta$.

\section{Transition Systems}

In this section we present an alternative approach to investigate universal
automata.	
	
	In order to specify what we mean by `universal PDA' or 
	`universal grammar', we introduce the notion 
	of \emph{Transition Systems}, a generalisation of finite automata
	that models any kind of computational device having
	an internal \emph{state} that can be altered by the occurrence of 
	an \emph{action} during the course of a computation.

\bigskip
\noindent
	{\bf Definition 1: (Transition System)} \label{def:TS}
	
		A \emph{Transition System} is a quintuple $(S,\Sigma,\delta,S_0,S_F)$ with
		
		\begin{tabular}{ll}
			$S$  							& a set of states,\\
			$\Sigma$						& an alphabet of transitions\\
			$\delta \subseteq S\times (\Sigma\cup\{\ew\}) \times Q$
											& a transition relation\\
			$S_0\ss S$			& a set of initial states\\
			$S_F\ss S$			& a set of final states\\
		\end{tabular}
		
		We write $s\stackrel{t}{\raa}s'$ for $(s,t,s')\in \delta$.
		A transition system is called \emph{finite} if $S\cup T$ is finite.
		The transition relation in a transition system can be extended 
		to finite sequences of transitions:
		\begin{itemize}
		 		 \item $s \stackrel{\ew}{\raa} s$ for all states $s$
		 					and the empty sequence $\ew$.
		 		 \item $s \stackrel{wt}{\raa} s'$ iff a state $s''$ exists,
		 					such that $s \stackrel{w}{\raa} s'' \land 
		 					s'' \stackrel{t}{\raa} s'$.
		\end{itemize}
		Let $\stackrel{*}{\raa}$ denote the transitive and reflexive closure
		of $\raa$.

		The \emph{language} of the transition system $A$ is
		$$L(A)= \{w\in\Sigma^*| s_0\stackrel{w}{\raa}s_f, s_0\in S_0, s_f\in S_F\}.$$

	Any finite automaton $A=(Q,\Sigma,\delta,q_0,Q_F)$ is also
	a transition system by definition. However, in a transition system
	$Q$ and $Q_F$ are generally not finite.
	Any PDA $A=(Q,\Sigma,\Delta,\delta,q_0,Q_F)$ defines a transition system
	whose states are all possible configurations $c\in Q\times\Delta^*$ of 
	the PDA, and transitions are defined by
	$(q,wx)\stackrel{a}{\raa}(q',wy) \iff (q,a,x,y,q')\in \delta$.

\bigskip

\noindent
{\bf Definition 2: (Universal Transition System)}

		  Let ${\bf X}\subseteq {\bf RE}$ be a language class below {\bf RE}.
		  The TS $A=(S,\Sigma,\delta,S_0,S_F)$ is called 
		  \emph{{\bf X}-universal} iff for any language 
		  $L\in \Sigma^*$ in ${\bf X}$ there is a state $s_L\in S$ such that 
								$L((S,\Sigma,\delta,s_L,S_F))=L$.

Any computing device that unambiguously defines a transition system will be 
considered {\bf X}-universal if its transition system is. 

Obviously there \emph{is} a {\bf REG}-universal
system, namely the disjoint union of all the possible NFA.
The interesting question is however, whether or not such a 
{\bf REG}-universal system can be defined by an NFA.
The following lemma recovers a theorem from \cite{Kud}
from a new perspective.

\bigskip

\noindent
{\bf Lemma 1:}
		  There is no {\bf REG}-universal NFA.

\noindent
{\it Proof :}
	For arbitrary $n\in \Nat,a\in \Sigma$ $L_n=\{a^n\} \in {\bf REG}$.
	Any {\bf REG}-universal system $U$ must have a state $s_{L_n}$ 
	from that on exactly $n$ steps can be made.
	If the set of states in $U$ was finite, this state
	could not exist for $n>|S|$, so $U$ must have an infinite set of states.
	Since any NFA defines a finite transition system, no {\bf REG}-universal 
	system can be defined by an NFA, and therefore no {\bf REG}-universal
	NFA exists.
\hfill$\Box$	

\section{Outlook}

We have shown that there doesn't exist
universal 1-way or 2-way finite automata nor pushdown
automata if encoding and decoding have to be done by
deterministic finite state transducers. However, if
2-way deterministic pushdown transducers are allowed and
for encoding
of a repetition of the specific pushdown automaton to
be simulated, depending on the length of input,
a universal pushdown automaton can be constructed.
Further research has to be done on encoding and decoding
device. Our conjecture is that 1-way deterministic are
not sufficient for the existence of a universal
pushdown automaton.

\section{Acknowledgement}
The author thanks Georg Zetzsche for many fruitful
discussions and contributions to this article.

\end{document}